\documentclass[%
 reprint,
 amsmath,amssymb,
 aps,
]{revtex4-1}

\usepackage{graphicx}
\usepackage{dcolumn}
\usepackage{bm}

\usepackage{graphics}
\usepackage{dcolumn}
\usepackage{bm}
\usepackage[caption=false]{subfig} 
\usepackage{color}
\usepackage{tikz}
\usepackage{subfig}
\usepackage[titles,subfigure]{tocloft}
\usepackage{soul}
\usepackage{scrextend}
\usepackage{hyperref}
\usepackage{footmisc}
\usepackage[titletoc,toc,title]{appendix}
\usepackage{dsfont}
\usepackage[normalem]{ulem}
\usepackage{mathtools}

\newcommand{\beq}{\begin{eqnarray}}
\newcommand{\eeq}{\end{eqnarray}}
\newcommand{\dg}{\dagger}
\newcommand{\bpm}{\begin{pmatrix}}
\newcommand{\epm}{\end{pmatrix}}

\newcommand{\ket}[1]{| #1 \rangle}
\newcommand{\bra}[1]{\langle #1 |}

\definecolor{purple}{rgb}{0.5,0,0.5}
\definecolor{dkgreen}{rgb}{0,0.5,0}

\newcommand{\pcsadd}{Center for Theoretical Physics of Complex Systems, Institute for Basic Science (IBS), Daejeon 34126, Republic of Korea}

\begin{document}

\title{Anomalous transport a in topological Wannier-Stark ladder}

\author{Kun Woo Kim}
\affiliation{\pcsadd}
\affiliation{Institut f\"or Theoretische Physik, Universit\"at zu K\"oln, 50937 Cologne, Germany}
\author{Alexei Andreanov}
\affiliation{\pcsadd}
\author{Sergej Flach}
\affiliation{\pcsadd}

\date{\today}

\begin{abstract}
    A dc (e.g. electric) field with commensurate lattice direction turns a single particle band structure in $d=3$ dimensions into an infinite set of equally spaced irreducible $(d-1)=2$-dimensional Wannier-Stark (WS) band structures that are spatially localized along the field direction. Particle transport is expected to be suppressed once the {WS} bands are gapped in energy. The topological character of the irreducible band structure leads to one-dimensional sets of {boundary} states which fill the energy gaps. As a result, eigenmodes are smoothly connected in energy and space and yield anomalous particle transport throughout the ladder. The number of {chiral boundary modes}  can be tuned by the dc field strength and manifests {through the distribution of dissipated energy and spatial motion, and the temperature dependence of angular momentum carried by particles.}
\end{abstract}

\maketitle

\section{Introduction}

Particle transport in quantum mechanical systems is of fundamental interest in condensed matter physics. When exposed to two thermalized reservoirs with different chemical potentials, particles are transferred due to incoherent energy relaxation.
For systems with almost perfect conductance such as Dirac semimetals and boundary modes in topological insulators, energy relaxation is mostly confined to the reservoir contacts~\cite{kawano2003spatial,grosse2011nanoscale}. 
If the chemical potential difference between two reservoirs is increased to values such that the potential drop between neighboring lattice sites is larger than $\hbar/\tau_{inc}$ where $\tau_{inc}$ is a characteristic inelastic scattering time, a Wannier-Stark (WS) ladder begins to develop (for a review, see~\cite{gluck2002wannier,kolovsky2004bloch}). 








The Stark effect is well known in the study of atomic energy splitting by external electric fields. When a strong dc electric field is applied to a lattice system, its electronic band structure shows a similar splitting. A dc (e.g. electric) field with commensurate lattice direction, e.g. parallel to any lattice vector of finite length, turns a single particle band structure in $d$ dimensions into an infinite set of equally spaced {irreducible}  $(d-1)$-dimensional Wannier-Stark band structures~\cite{wannier1960wave}. 
With the advance of experimental techniques both in superlattice semiconductors~\cite{mendez1988stark} and cold atoms in optical lattices~\cite{wilkinson1996observation, kolovsky2004bloch},  photonic lattices~\cite{mukherjee2015modulation}, and bulk GaAs using transient bias technique~\cite{schmidt2018signatures}, the WS ladder spectrum {has been well confirmed experimentally}. WS systems contain the physics of strong electric fields under non-equilibrium conditions~\cite{lee2014dielectric,li2015electric}, strongly localized states in space~\cite{kolovsky2018topological}, disorder and correlation effects~\cite{schulz2019stark, van2019bloch}, topological characters associated with multiple bands~\cite{lee2015direct,kim2016surface,kim2019floquet} and stay in the focus of current research~\cite{sie2016observation,kruchinin2018colloquium,jin2018high, bosco2019thermoelectrically}.

WS Hamiltonians are closely related to time-periodic Floquet Hamiltonians, $H_{\text{Floq}}(t+T) = H_{\text{Floq}}(t)$. A static external electric-field $\vec F$ can be gauged into a wave {vector:} $\vec k \rightarrow \vec k - \vec F t$. With discrete translational symmetry it follows $H(k) = H(k+2\pi/a)$ ({$a$ is a lattice constant;} we drop the vector notation for convenience) {and} the gauged WS Hamiltonian $H(k- Ft)$ is Floquet time-periodic with period {$T = 2\pi/(F a)$}. {The first attempt to make use of the Floquet formalism to obtain  Wannier-Stark states was reported by Gl{\"u}ck et al.~\cite{gluck1998calculation, gluck2000resonant,gluck2002coherent,wacker2013nonequilibrium}. The connection between a single particle d-dimensional WS Hamiltonian  and a $(d-1)$-dimensional Floquet Hamiltonian was later studied with concrete examples carrying nontrivial topological characters in Bloch bands~\cite{gomez2013floquet,kim2019floquet}}. Note however that the WS wave functions live in the whole physical $d$-dimensional space, at variance {with} Floquet systems where wavefunction are confined to a $(d-1)$-dimensional physical space. 

{Recent research of Floquet physics focused on constructing topological quantum systems~\cite{kita2010topo,lind2011floq,
rudner2013anomalous,rudner2019floquet,rudner2020the}. At the same time Wannier-Stark research focuses on computing particle currents under strong electric fields as a way to observe the Wannier-Stark ladder~\cite{lee2002nonequilibrium,lee2006quantum}.} In this work, we present a study of edge particle and thermal transport along the dc electric field direction of a Wannier-Stark ladder with a nontrivial topological character in $d$=3 space dimensions. Below a certain electric field strength one-dimensional boundary modes are generically connecting energy-spaced Wannier-Stark bands. The boundary mode number is controlled by the electric field strength. Thanks to coupling to an incoherent scattering source, particles can propagate along the Wannier-Stark ladder as they relax energy to the bath.

{As opposed to in-gap impurity states for which spatially localized modes have random energies and spatial locations, the topological protection of boundary states ensures continuous distribution of eigenmodes in energy space with a finite spatial overlap among neighboring modes. This therefore provides a weak but robust transport of particles across the energy gap regardless of the cutoff energy of heat bath, which otherwise will show transport properties similar to insulators.
}




\begin{figure*}
\begin{center}
    \includegraphics[width=2\columnwidth]{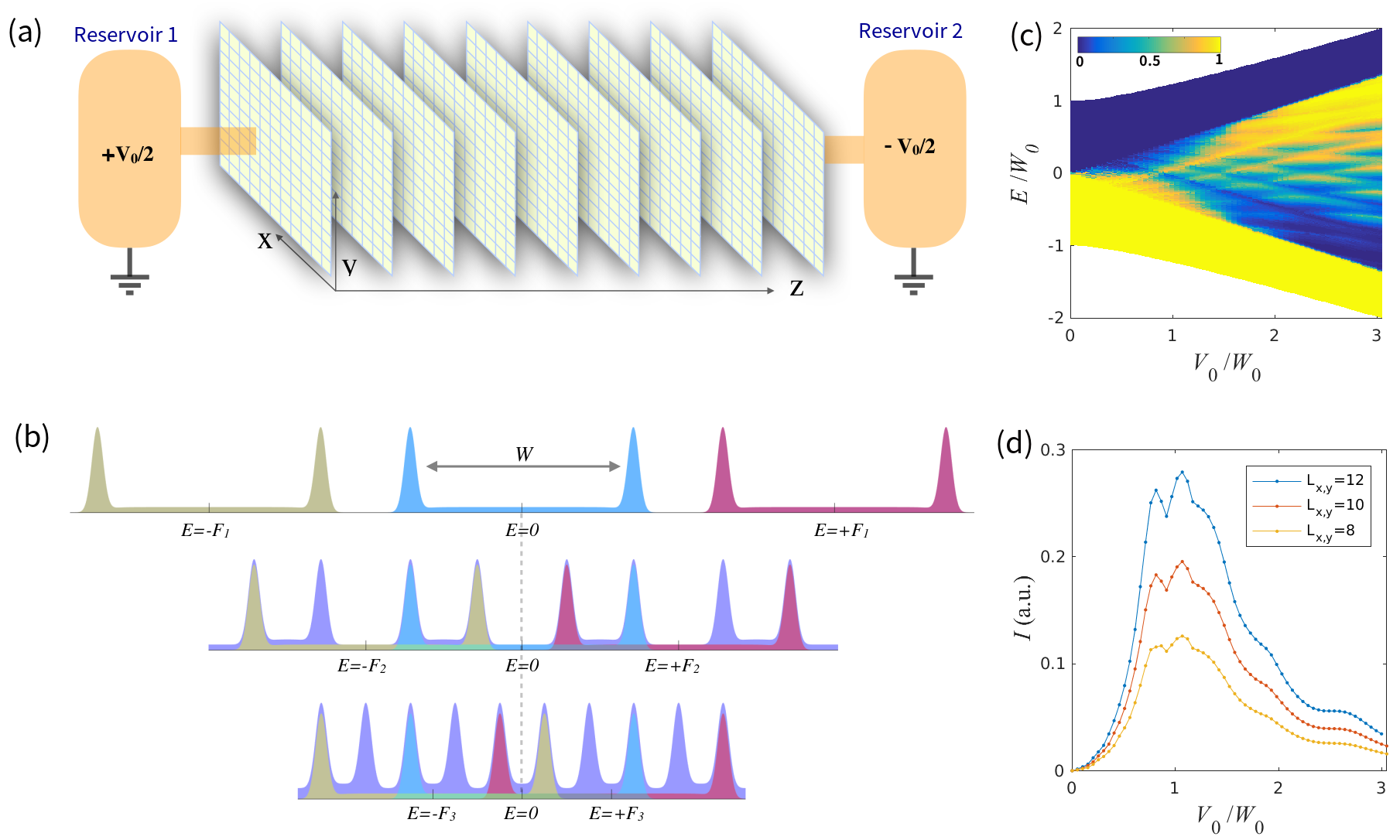}
    \caption{{Sketch of the topological Wannier-Stark ladder}. (\textbf{a}) Between two reservoirs with chemical potential difference $ V_0$,  2D Chern insulators are connected in the $\hat z$-direction, along which  the system  carries a steady particle current.  (\textbf{b})  Sketch of band inversion between the Wannier-Stark bands. A strong electric field $F_1 (>W)$ separates the  energy spectrum of each 2D layer, with 1D boundary modes filling the energy gap of a pair of Chern bands. Upon lowering the external field strength ($F_2<W$), band inversions take place, followed by doubling the  of boundary modes at $E=\pm F_2/2$. Further decreasing the electric field ($F_3<W/2$), the of boundary modes is tripled at $E=0, \pm F_3$. (\textbf{c}) Occupation number of particles as a function of voltage drop $V_0$ divided by a bandwidth of the 3D lattice model without bias. (\textbf{d}) Particle current between two reservoirs in steady states with increasing voltage drop at different cross section $L_{x}$=$L_y$=8, 10, 12 and $L_z$=9.}
    \label{fig:1}
\end{center}
\end{figure*}

\hfill

\section{Topological Wannier-Stark ladder}

Figure~\ref{fig:1} provides a schematic understanding of the energy spectrum of the topological Wannier-Stark ladder (TWSL).
A 3D lattice model composed of layers of a 2D Chern insulator with chiral boundary modes circling on the open boundary. When a strong external electric field {$F=eV_0/L_z$} is applied along the $z$-direction, the spectrum of the Chern insulators is identically repeated with every {$\Delta E = ea_zF$ (lattice constant $a_z$)}, forming a set of the Wannier-Stark bands extended in the $xy$-plane while localized in $z$-direction. 
The density of states within the energy gap $W$ of two WS bands is filled  with chiral boundary modes. In Fig.~\ref{fig:1}(b) the three sets of WS bands {near zero energy} are colored to enhance visualization. With decreasing electric field strength, the first band inversion takes place at $E=\pm W/2$, and is followed by a stepwise increment of the DoS inside the bulk energy gap. With further decreasing of the field strength ($F=1.4W, F=0.7W, F=0.4W$), energy gaps are flooded with additional sets of boundary modes. Their maximum number is limited {by the ratio of the energy gap $W$ and the intrinsic band width, and other inelastic scattering sources broadening the spectrum}. 
With the set of chiral boundary modes connecting Wannier-Stark bands, our main question concerns their role in particle transport from one particle reservoir to the other and {the related thermal energy emission in the course}. Figures~\ref{fig:1}(c), d show the distribution of particle occupancy and the particle current between the two reservoirs in a steady state as the external electric field is tuned, respectively. The details of the calculations will be explained in the following section. The emergence of the Wannier-Stark ladder {is shown} with increasing voltage drop $V_0$ normalized by a bandwidth $W_0$ of 3D lattice model without bias {($F=0$)}. The distribution begins to show a dramatic deviation from the Fermi-Dirac one at $V_0/W_0=1$, and at the same time the differential conductance {$dI/dV$ turns} negative, signifying deviation from the transport of conventional conductors. \newline

\section{Model Hamiltonian and Pauli master equation}.
{To be specific}, we employ a model {tight-binding} Hamiltonian of the Floquet topological insulator {on a cubic lattice with two states per site}~\cite{rudner2013anomalous}:
\beq
    \hat H _{WS} = \sum_n\left[ \vec{d}\cdot\vec{\sigma}-F n\right] c^\dg_nc_n+\frac{\Delta}{2} \sigma_3 (c^\dg_{n+1}c_n + c^\dg_{n}c_{n+1}),\nonumber\\ \label{H}
\eeq
where $d_1= \alpha \sin k_x$, $d_2= \alpha \sin k_y$, $d_3=\mu-J - 2\beta (2-\cos k_x- \cos k_y) + J\cos k_x \cos k_y )$ and $\vec \sigma = (\sigma_1,\sigma_2,\sigma_3)$ are the Pauli matrices. The lattice spacings $a_x=a_y=a_z=1$ and $e=1$. The hopping strength between layers $\Delta=10$, and the intralayer hopping strengths $\mu=3$, $\alpha=4$, $J=\beta=1.4$, {are chosen such that each 2D WS band carries the Chern number $\pm1$ in the large field limit, $F/W_0 \gg 1$. The index $n$ {labels the 2D layers} with $H_{2D}=\vec d\cdot \vec \sigma$ in the $xy$-plane, {as shown in Fig.~\ref{fig:1}(a)}. The lattice translation symmetry is broken in the $z$-direction due to the presence of the electric field $F$, which adds a {stepwise increase of potential energy} to the 2D layers. {Instead}, the model acquires a combined symmetry $\hat T_{ZE}$ of discrete translation and energy-shift in the $z$-direction $\{n\rightarrow n+1, E \rightarrow E-F\}$, as observed in the energy spectrum of eigenmodes {in Fig.~\ref{fig:2}(a)}. 



We assume a finite extension of the system in the $z$ direction with $-L_z/2 \leq z \leq L_z/2$ and sizes $L_x,L_y$ in the $x$ and $y$ directions respectively. Particles at chemical potential $\mu_1=V_0/2$ are released from the {reservoir $1$ at $z=L_z/2$}. For them to reach the reservoir $2$ with $\mu_2=-V_0/2$ {at $z=-L_z/2$}, a corresponding energy difference $\sim eV_0$ must be released or dissipated. We {add} an incoherent scattering source (for example, phonons) with a well defined temperature interacting with the fermionic particles in the system. To maintain the integrity of the Wannier-Stark ladder, we assume the coupling strength to be small and the maximum energy carried by one phonon ($\hbar \omega_D$, the Debye frequency) to be smaller than the potential difference between neighboring 2D layers, $F$. We will discuss below the impact of variations $\omega_D$ on the transport properties of TWSL. The Pauli master equation, which is valid for the calculation of steady states~\cite{fischetti1998theory,rott2002field}, is employed to compute the occupation numbers $f_n\coloneqq f(\epsilon_n)$ of eigenmodes $n$: 
\beq
    \frac{d f_n}{dt} &=&     \sum_{m\neq n} P_{nm},\nonumber\\
     &=& \sum_{m\neq n} W_{nm} (1-f_n) f_m - W_{mn}  (1-f_m)f_n,
    \label{pme}
\eeq
where $P_{nm}$ is the scattering rate from eigenmode $m$ to $n$. The scattering strength $W_{nm}$ is determined by the density overlap between two eigenmodes,  the temperature of phonon bath, and the phonon density of state (see {Appendix A} below for details).


Figure~\ref{fig:1}(d) shows the rate of particle number going into the reservoir $2$ from the TWSL as a function of the electric field strength for three cross-section sizes ($L_x\times L_y$) 
with $L_z=9$ in steady states, which is equal to the rate of particle number getting out of the reservoir $1$ (see {Appendix A} for the setup of reservoirs). {At a small voltage drop $V_0= FL_z<W_0$}, the current is increasing with $V_0$ as more active transport channels become available. The slope $dI/dV$ is not constant since the density of states {depends on energy}.  On the other hand, when the field strength is further increased, the current decreases with oscillations reflecting the series of  {energy gap closings} {in the} WS bands. The occurrence of the Wannier-Stark ladder is visible in the map of occupation numbers in Fig.~\ref{fig:1}(c) through the sequence of occupied and unoccupied energy bands. \newline


\begin{figure}
\begin{center}
    \includegraphics[width=1\columnwidth]{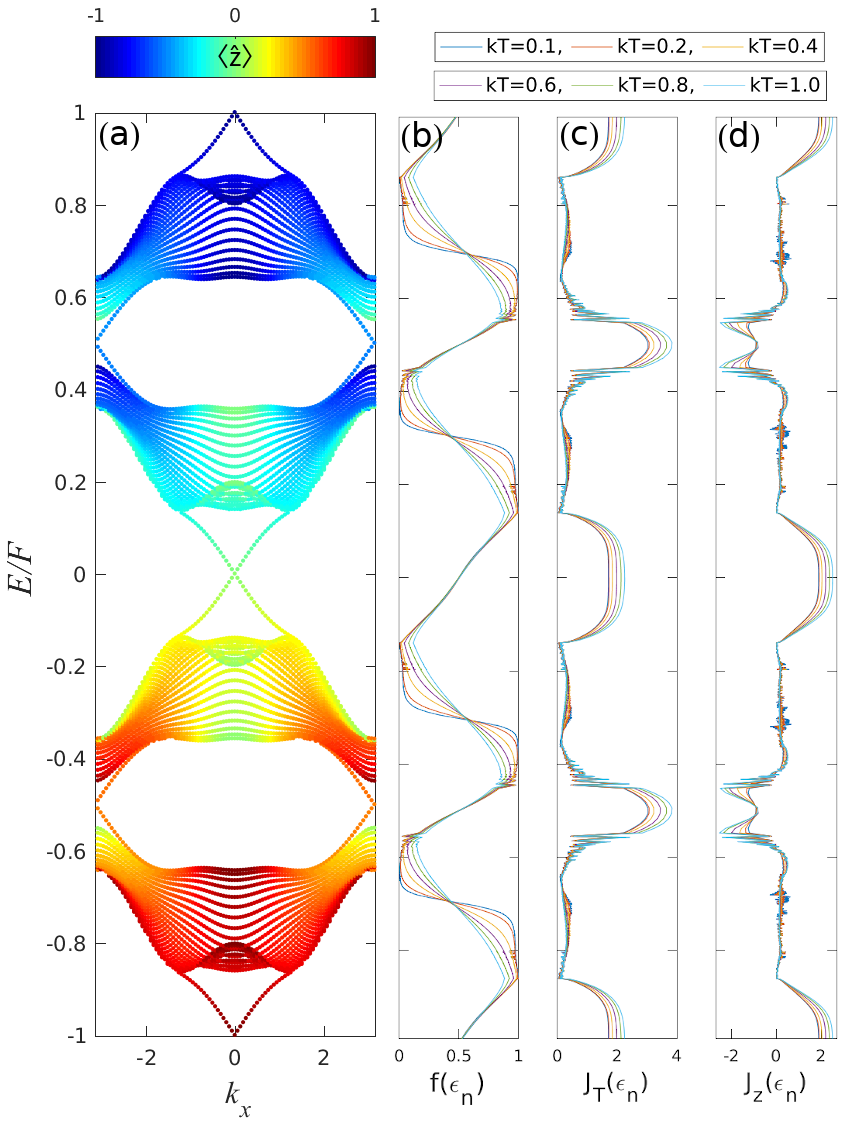}
    \caption{{Energy-momentum dispersion relation and nonequilibrium steady states in the TWSL}.  (\textbf{a})  $E(k_x)$  at $F=15$. 
The color code indicates the average position $\hat z$ of each eigenmode. (\textbf{b}) Occupation number $f(\epsilon_n)$ in steady states at temperature $kT=0.1-1.0${, filling fraction $\nu=0.5$.} (\textbf{c}) Energy dissipation rate $J_T(\epsilon_n)$ associated with each eigenmode $n$. (\textbf{d}) Spatial current $J_z(\epsilon_n)$ in the direction of the external field. For the calculation of steady states, we use $N_x$=$N_y$=22 with open boundary conditions. }
    \label{fig:2}
\end{center}
\end{figure}

\section{Nonequilibrium steady states}
\noindent Using the model Hamiltonian and the Pauli master equation, in this section we present steady state results {for} occupation number $f(\epsilon_n)$, energy dissipation rate $J_T(\epsilon_n)$, and spatial currents $J_z(\epsilon_n)$ associated with each eigenmode {of} TWSL. {These} three quantities show distinct behavior {in the bulk and chiral boundary modes}, signifying their crucial role in the particle transport of the TWSL {(see Appendix B for a comparison to in-gap impurity states ). } 

\begin{figure*}
\begin{center}
    \includegraphics[width=1.9\columnwidth]{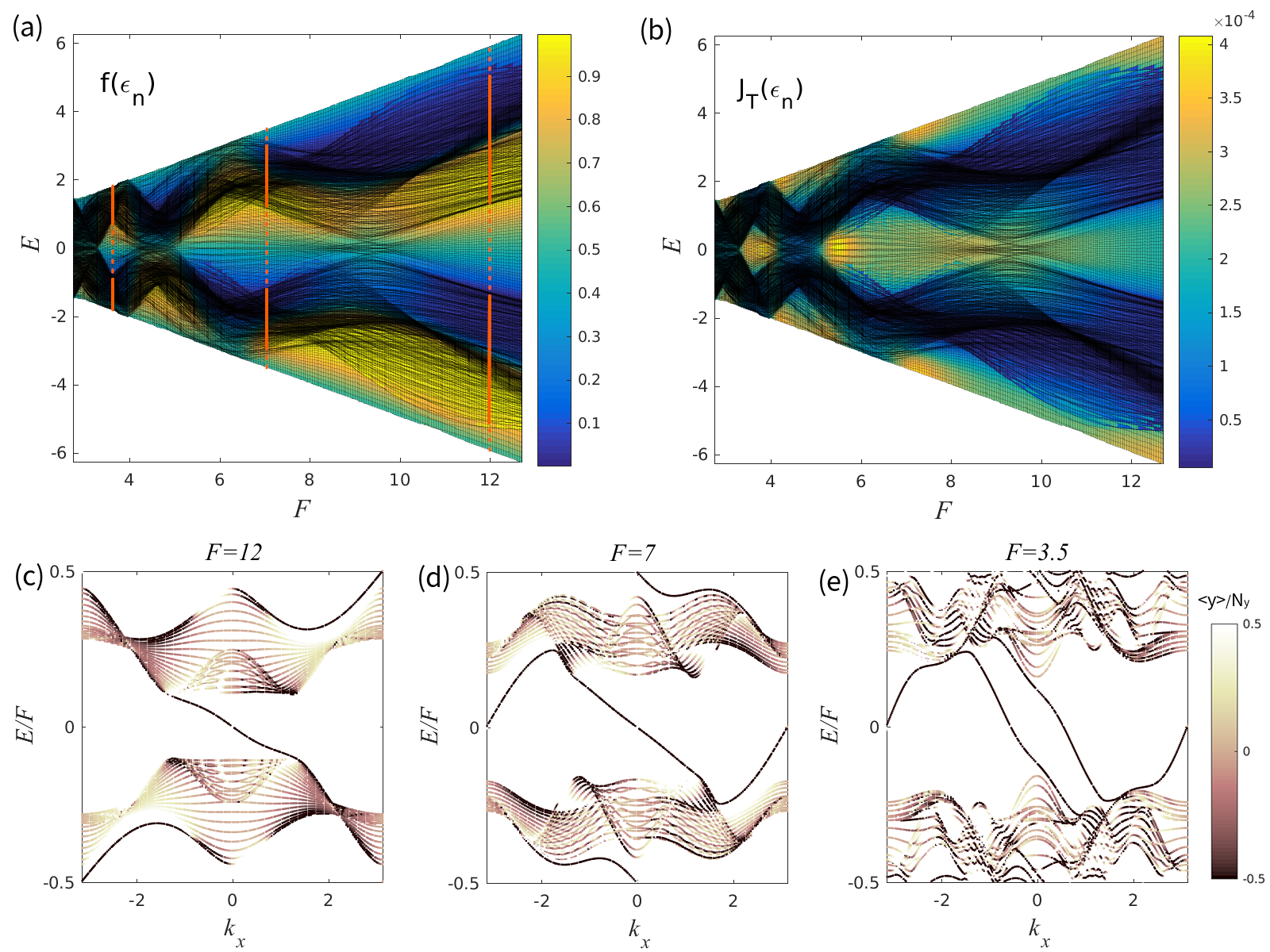}
    \caption{{Energy resolved occupation numbers and energy dissipation rate}. (\textbf{a}) The occupation numbers $f(\epsilon_n)$ of particles in eigenmode $n$, within one irreducible {set of} WS band in $E\in[-F/2,F/2]$ as a function of the external field strength $F$. {Thin black lines indicate the trace of eigenenergy in TWSL as a function of $F$. For three values of the external field strength, $F$=3.5, 7, and 12, the energy windows of the WS bulk modes are marked with thick solid vertical lines, and the ones of the chiral boundary modes with dotted lines. (\textbf{b}) Energy dissipation rate $J_T(\epsilon_n)$ from scatterings associated with eigenmode $n$. (\textbf{c-e}) Momentum-energy dispersion relation $E(k_x)$ for three representative electric field values $F=12, 7, 3.5$ with a sizable energy gap. The color coding is chosen to emphasize eigenmodes near the open surface at $y=-L_y/2$ {(black)}.}}
    \label{fig:3}
\end{center}
\end{figure*}

Figure~\ref{fig:2}(a) shows the dispersion relation of the TWSL for $F=15$.
Here, the periodic boundary condition is assumed along the $x$-direction in order to illustrate the chiral boundary mode dispersion in momentum $k_x$. With open boundaries at $y = \pm L_y/2$, two boundary modes with opposite chiralities appear in pair. The color code indicates the expectation value $\langle \hat z \rangle$ of the $z$-position of the eigenmodes. While the overall position $\langle \hat z \rangle$ decreases with energy {since so does} the potential energy $V_{\text{pot}} = - Fz$, at energy $E=F/2$ where boundary modes appear as a result of the band inversion the position $\langle \hat z \rangle$ increases with energy.
Using the set of eigenmodes of the Wannier-Stark system, we compute particle and energy transport characteristics {by solving} the Pauli master equation~\eqref{pme}. Due to the presence of the symmetry $\hat T_{ZE}$ we pick one irreducible eigenmode set{, $-F/2\leq E\leq F/2$,} and add periodic boundary conditions in the energy domain. This allows us to compute the steady nonequilibrium state of the system for a region with energy $E$ away from the reservoirs, e.g. $|E-V_0|/F \gg 1$. For the {setup} of Fig.~\ref{fig:2}(b-d), we take the number of particles  to fill one half of the eigenmodes in the irreducible TWSL band: $\nu = \sum_n f(\epsilon_n)/(2N_xN_y) = 0.5$.

Figure~\ref{fig:2}(b) shows the steady state occupation numbers of the WS eigenmodes {for} temperatures $kT=0.1-1$ with $w_D/F=0.02$. Within the bulk bands, particles can efficiently relax their energy as there are {roughly} $\sim 2L_x L_y$ scattering channels. On the other side, the boundary  modes between two WS bands have only few scattering channels at hand. Therefore, the occupation number is nearly unity at the bottom of each WS band  where particles are passing through a bottleneck to enter boundary channels. Then, particles are transported down to the top part of the next WS bulk band where the occupation is close to zero. {Since the Debye frequency is much smaller than the energy gaps between the WS bands}, the interband particle transfer occurs predominantly via boundary modes. 
Note that with varying temperature the particle occupation of bulk modes follows {closely} the Fermi-Dirac distribution, $f(\epsilon_n) \simeq 1/(e^{(\epsilon_n-\mu_i)/kT}+1)$ with the WS band {dependent}  chemical potential $\mu_i$, while the occupancy of boundary modes is essentially independent of the temperature, and {does not thermalize}. As a consequence, {certain physical observables such as the angular momentum discussed below show a non-trivial temperature dependence}.

\begin{figure*}
\begin{center}
    \includegraphics[width=2\columnwidth]{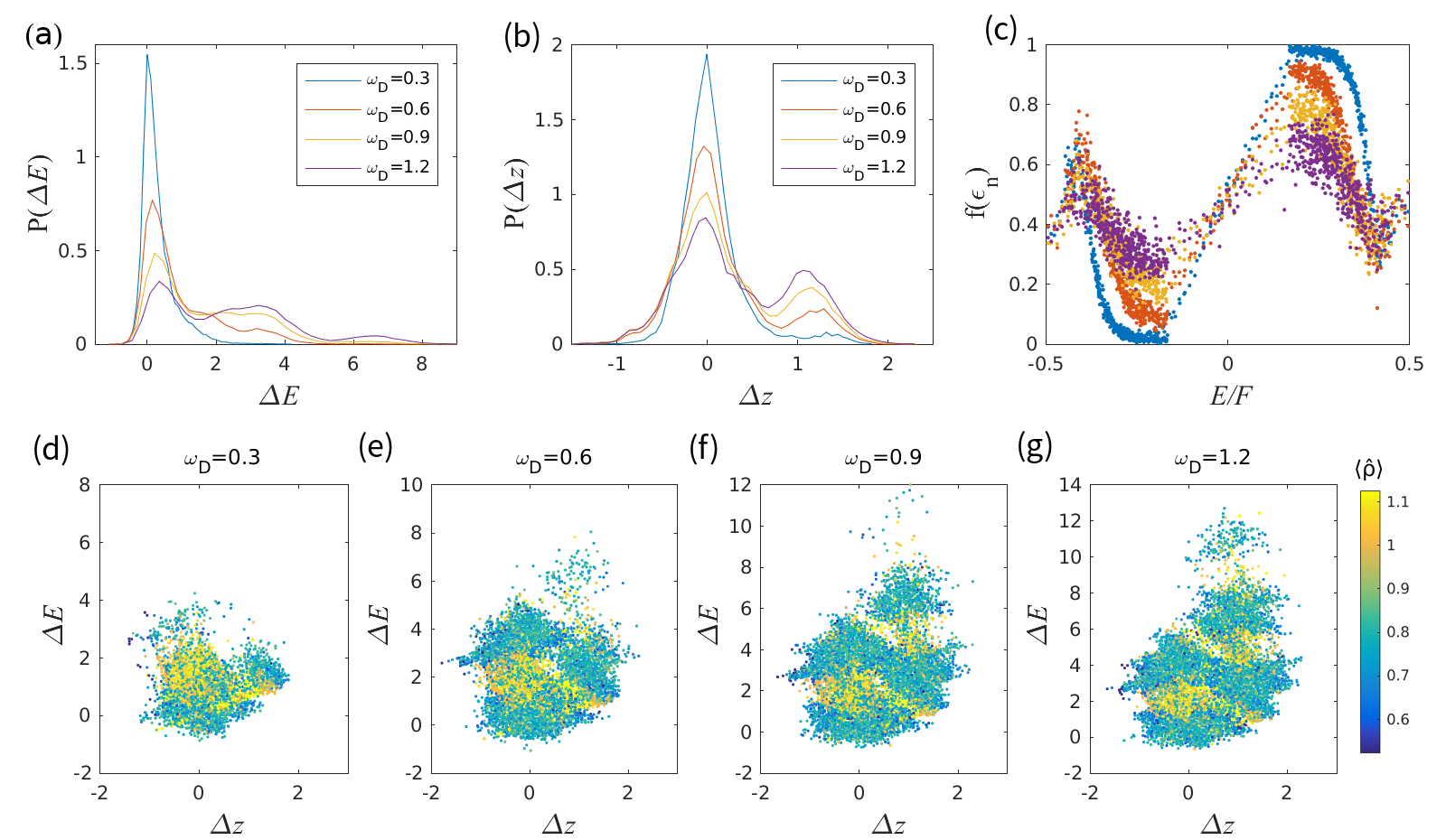}
    \caption{{Dissipated energy and spatial displacements of particles in TWSL.} (\textbf{a-b}) Distribution of dissipated energy and spatial displacements of particles along $\hat z$ at $F=7$, $kT=0.1$ for different Debye frequencies: $\omega_D=0.3, 0.6, 0.9, 1.2$. (\textbf{c}) Occupation number of particles $f(\epsilon_n)$ for the {same set of Debye frequencies (the same color code)}.  (\textbf{d-g}) Correlations of dissipated energy and spatial displacements of particles. Color indicates the expectation value of $\hat \rho =\left[ (2\hat x/L_x)^2 + (2\hat y/L_y)^2\right]^{1/2}$ of eigenmodes associated with scatterings, distinguishing boundary modes from the WS bulk modes: scatterings marked by yellow take place near to {the surface.}}
    \label{fig:4}
\end{center}
\end{figure*}

Figure~\ref{fig:2}(c) shows the {energy dissipation rate associated with WS eigenmodes:}

\beq
    J_T(\epsilon_n ) = -\sum_{m\neq n} (\epsilon_n - \epsilon_m) P_{nm},
\eeq
where the scattering rate between eigenmodes is weighted by the amount of released energy. $J_T(\epsilon_n)$ is always positive since  the scattering events of emitting phonons are more probable than those of absorbing a phonon, $P_{nm}>0$ for {$\epsilon_n < \epsilon_m$} (see {Appendix A}).
The energy disspation rate per eigenmode $J_T(\epsilon_n)$ is largest at $E=0$ and $E=\pm F/2$ where the particle distribution is far from the equilibrium, while it is suppressed within {WS} bands which are well thermalized as reflected through the Fermi-Dirac distribution of their occupancies.

In addition to energy dissipation, the particle transport in the TWSL yields spatial displacements in the direction of the external field. In Fig.~\ref{fig:2}(d), the spatial current of particles associated with each eigenmode is computed:
\beq
    J_z(\epsilon_n ) = \sum_{m\neq n} (z_n - z_m) P_{nm},
\eeq
where $z_{n} = \bra{n}\hat z \ket{n}$. The spatial current $J_z(\epsilon_n )$ is again maximum at zero energy in the chiral boundary modes, while it shows a reversed motion with negative sign at $E=\pm F/2=\pm 7.5$ reflecting the presence of inverted {WS} bands. 
As it is clearly seen in Fig.~\ref{fig:2}(c-d), both the energy dissipation rate and spatial current show a certain correlation in their magnitudes, which will be further {discussed} in the section on physical observables.\newline



\section{Topological phase transition in TWSL}
As shown in Fig.~\ref{fig:1}(b), a WS band inversion takes places repeatedly with the decrease of electric field strength $F$. Figure~\ref{fig:3}(a) shows numerical results for the occupation number $f(\epsilon_n)$ of eigenmode $n$ as a function of energy $E$ and field strength $F$. The eigenenergies {as functions of $F$} are plotted with thin solid lines indicating the location of the  Wannier-Stark bands and the edge/boundary modes which fill into the energy gap with a relatively larger energy spacing. For three values of the electric field strength $F=12, 7, \text{and } 3.5$ we show the dispersion relations in Fig.~\ref{fig:3}(c-e) with the expectation value of $\hat y$ in the color coding to emphasize eigenmodes {localized near the open surface} at $y=-L_y/2$. The above cases demonstrate that the number of 1D edge/boundary modes localized at the open boundary  at $E=0$ and $E=\pm F/2$ are tuned by the external electric field strength. 
The boundary modes are connecting fully occupied bulk modes with $f(\epsilon_n)\simeq 1$ (yellow) to almost empty bulk modes $f(\epsilon_n)\simeq 0$ (dark blue) in Fig.~\ref{fig:3}(a). {By releasing energy through scattering off phonons,}
particles \textit{slide} down through boundary modes in energy between two Wannier-Stark bands. With a steep gradient of occupation number and only a few channels to scatter into, the {rate of thermal energy emission $J_T(\epsilon_n)$} is particularly enhanced at $E=0$ and $E=\pm F/2$ as {shown} in Fig.~\ref{fig:3}(b). The two sets of figures show the topological phase transition with external electric field $F$, accompanied by the change of particle occupation in steady states and the thermal energy emission signified by the presence of boundary modes. 
\begin{figure*}
\begin{center}
    \includegraphics[width=2\columnwidth]{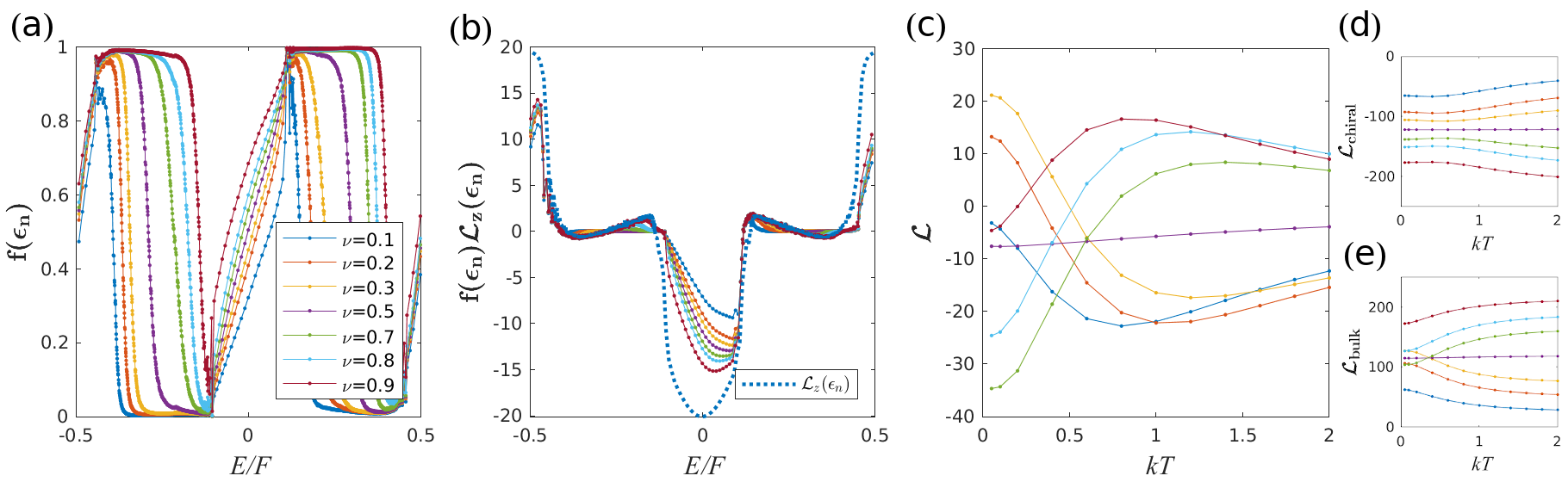}
    \caption{{Angular momentum at different particle fillings and temperature in TWSL.} (\textbf{a}) Occupation number of particles $f(\epsilon_n)$ at $F=15$, $kT=0.1$, and filling fraction $\nu=0.1 - 0.9$ (legend). (\textbf{b}) Angular momentum carried by particles sitting in eigenmode $n$, $f(\epsilon_n)\mathcal L_z(\epsilon_n)$, is plotted in the same set of filling fractions. As a guideline, the angular momentum $\mathcal L_z(\epsilon_n)$  is plotted with dotted line. (\textbf{c}) The sum of angular momentum $\mathcal L =\sum_n f(\epsilon_n)\mathcal L_z(\epsilon_n)$ as a function of temperature for different filling fractions.  (\textbf{d-e}) Angular momentum carried by chiral boundary modes $\mathcal L_{\text{chiral}}$ and the WS bulk modes $\mathcal L_{\text{bulk}}$ are  plotted as a function of temperature, $\mathcal L=\mathcal L_{\text{chiral}} + \mathcal L_{\text{bulk}}$.}
    \label{fig:5}
\end{center}
\end{figure*}

\section{Experimental observables}

\noindent In the previous section, we showed the comprehensive maps of the occupation number of particles in the topological Wannier-Stark ladder and the map of energy dissipation rate of eigenmodes. As the external electric field $F$ is tuned, the number of boundary modes within the energy gap {varies} as well as the Chern invariant of {the} {WS} bulk bands. While the two quantities, $f(\epsilon_n)$ and $J_T(\epsilon_n)$, provide useful microscopic information of the TWSL, eigenenergy-resolved quantities are hardly accessible in experiment. Instead, their energy-integrated statistics can be measured. This section {discusses} the following physical observables {related to the} boundary modes in the {TWSL}: (i) the distribution of dissipated energy, (ii) the distribution of spatial displacements of particles in transport direction, (iii)  the sum of angular momentum carried by particles as a function of temperature and filling fraction.

\subsection{Distribution of dissipated energy and spatial displacements}
Figure~\ref{fig:4}(a-b) shows the {distribution} of dissipated energy $\Delta E$ and spatial displacements of particles along the $z$-direction $\Delta z$ in the steady state of the TWSL {for} several Debye frequencies $\omega_D=0.3, 0.6, 0.9, 1.2$. 
\beq
P(\Delta E) &=& \sum_{m,n} P_{nm}\delta( \epsilon_n- \epsilon_m + \Delta E ),\\
P(\Delta z) &=& \sum_{m,n} P_{nm}\delta( z_n- z_m -\Delta z ),
\eeq
where the sum {runs over the eigenmodes of} TWSL. To reduce the finite size effect, $\delta(x)\simeq \pi^{-1}\eta/(x^2+\eta^2)$ with $\eta=0.1$ is used. 
External field strength $F=7$ is chosen without loss of generality. For both plots, the {main peaks are} located at {$\Delta E = \Delta z = 0$}, {and appear due to particles of the bulk modes} with occupation close to unity. The {distribution} is markedly asymmetric reflecting the fact that particles are moving to {lower energies and positions $z$}. The {broad secondary peaks following the main one} are caused by inter-band scattering, whose rate increases with {the increase of the Debye frequency $\omega_D$}. In the presence of boundary modes connecting {neighboring WS bands}, no matter how small a Debye frequency $\omega_D$ is, particles can always find their paths to lower energy, and the dips between the {peaks} are filled by particle motions via boundary modes.

The particle occupancies {$f(\epsilon_n)$} for different $\omega_D$ are shown in Fig.~\ref{fig:4}(c). For small $\omega_D$ particles follow the Fermi-Dirac distribution within the WS bands with phonon temperature $kT=0.1$ (see Fig.~\ref{fig:3}(d) for the energy dispersion relation). However, with increasing $\omega_D$ this quasi-equilibrium is lost, since {more and more} particles are efficiently scattered to the next Wannier-Stark band via fast direct intraband energy relaxation.

{Figures~\ref{fig:4}(d-g) show} the correlation between dissipated energy {$\Delta E$} (vertical axis) and the spatial displacements of particles {$\Delta z$} (horizontal axis) for the same set of Debye frequencies, $\omega_D=0.3, 0.6, 0.9, 1.2$, for $10^5$ scattering events. The color of data points indicates the expectation value of $\hat \rho =\left[ (2\hat x/L_x)^2 + (2\hat y/L_y)^2\right]^{1/2}$ {within eigenmodes associated with scatterings, used as a measure distinguishing boundary modes from bulk ones}. Interestingly, the parameter space is divided by alternating scattering regions with different colors. If the boundary modes are absent, the regions of bulk scattering modes are disconnected owing to the energy gaps between WS bands. Thus, from the maps of correlations  one is able to identify the presence of boundary modes and their role in TWSL.

\subsection{Angular momentum}\label{SecAng}
Next, we turn our attention to the angular momentum carried by particles in the TWSL. Being localized on the open surface of a 3D lattice, the chiral motion of boundary modes generates a significant angular momentum along the $\hat z$-axis. Also the occupation of boundary modes is less sensitive to the temperature compared to the bulk modes, {as discussed earlier, see Fig.~\ref{fig:2}(b)}. This  provides an opportunity to characterize the presence of boundary modes in TWSL from the direct measurement of an angular momentum \cite{goldman2012detecting}, or local magnetic field if particles carry a charge~\cite{nowack2013imaging}. In the following numerical demonstration, the external field strength $F=15$ is chosen which corresponds to the {setup} of Fig.~\ref{fig:2}. 



The angular momentum in the direction of external field is associated with position and velocity operators: $ \mathcal{\hat L}_z = \hat x \hat v_y - \hat y \hat v_x$, where $\hat v_j = i[\hat r_j, \hat H]$. Thus,  the sum of angular momentum of particles in TWSL in a steady state is: 
\beq
   \mathcal  L &=& \sum_n f(\epsilon_n)L_z(\epsilon_n),
\eeq
where $\mathcal L_z(\epsilon_n) =i \bra{\psi_n} \hat x\hat H\hat y - \hat y \hat H\hat x \ket{\psi_n} $ is the angular momentum carried by eigenmode $n$. The sum runs over eigenmodes $n$ within an irreducible WS band structure around $E=0$. The occupancies of eigenmode $f(\epsilon_n)$ for different {filling fraction} $\nu=\sum_{n} f(\epsilon_n)/(2N_xN_y)=0.1-0.9$ are plotted in Fig.~\ref{fig:5}(a).  With increasing {filling fractions}, the local chemical potential $\mu_i$ of each WS band is increasing to maintain a quasi-equilibrium at $kT=0.1$.   At the same time, the occupancy of boundary modes stays  highly non-equilibrium  with  $\frac{df}{dE}>0$, as they are connecting the two neighboring WS bands.

 When the {TWSL} is completely filled, $f(\epsilon_n)=1$, the sum of  angular momentum $\sum_n\mathcal L_z(\epsilon_n)=\text{Tr} (\mathcal {\hat L}_z )=0$, because {we can always choose a basis, such that} $\ket{\phi_n}$ with $\bra{\phi_n}\mathcal {\hat L}_z \ket{\phi_n}=0$ for all $n$.
However, for  a filling fraction $\nu<1$, $\mathcal L$ is in general non-zero as the distribution of particle occupation is strikingly different for the WS bulk modes and  boundary modes as a function of temperature and filling fraction.  The eigenenergy-resolved angular momentum carried by particles, $f(\epsilon_n)\mathcal L_z(\epsilon_n)$, is shown in Fig.~\ref{fig:5}(b) {for} {the same set of filling fractions $\nu=0.1 - 0.9$}. Note that angular momenta  of chiral boundary modes around $E=0$ and $E=\pm F/2$ are distinctively large compared to that of the WS bulk modes, since the former is localized at the open surface with uni-directional group velocity.




Unlike systems in equilibrium where the role of temperature is often reduced to energy broadening and the diminishing of quantum effects, in TWSL the influence of  temperature on the angular momentum is dramatic. Figure~\ref{fig:5}(c) shows the temperature dependence of the sum of angular momentum at different fillings. {At half filling $\nu=0.5$, $\mathcal L$ is nearly insensitive to the change of temperature as a result of the symmetric angular momentum $\mathcal L_z(\epsilon_n)$ with respect to $E=0$ (see the dotted line in Fig.~\ref{fig:5}(b)), and the particle-hole symmetric occupation number $f(\epsilon_n)\simeq 1-f(-\epsilon_n)$ (see Fig.~\ref{fig:5}(a) and Fig.~\ref{fig:2}(b)).} 
As the filling is tuned away from the half, {$\mathcal L$ becomes sensitive to temperature and shows abrupt variations}. In Fig.~\ref{fig:5}(d-e), $\mathcal L$ gets contributions from the boundary modes $\mathcal L_{\text{chiral}}$ and bulk modes $\mathcal L_{\text{bulk}}$. Surprisingly, the temperature dependence of $\mathcal L_{\text{chiral}}$ and $\mathcal L_{\text{bulk}}$ for a given filling fraction marked by the same color shows opposite behavior, which is the reason for the non-trivial temperature dependence of the sum of  angular momentum carried by particles in TWSL. 
 \newline






\section{Discussion}
The Stark effect in a lattice, the Wannier-Stark ladder, was experimentally observed in semiconductor superlattice structures \cite{mendez1988stark} and cold atoms in optical lattices \cite{wilkinson1996observation}. Despite its continued interest, most experimental and theoretical studies are limited to one-dimensional lattices in a strong field. { In this manuscript, we theoretically study topological phases of a 3-dimensional Wannier-Stark ladder  which shares the same topological property with Floquet topological  phases in (2+1) dimensions. Moreover, the occupation number of particles in steady states is computed by solving the master equation which allows us to compute transport-related physical observables. 

The essential ingredient of our proposal, the topological Wannier-Stark ladder, leads to layers of two-dimensional topological bands and tto heir coupling in the direction of an applied external field. With  intensive  interest in topological matter, high quality two-dimensional topological bands are engineered both in solid state devices \cite{chang2013experimental} and in ultracold atomic setups \cite{aidelsburger2015measuring,wu2016realization,cooper2019topological}. 
Thus, the extension to an array of two-dimensional layers as described in Fig.~\ref{fig:1}(a) is currently within experimental reach. 
We propose to introduce a constant potential gradient across the two-dimensional coupled layers to induce the topological Wannier-Stark ladder accompanying the chiral boundary modes in every energy gap between WS bands (see Fig.~\ref{fig:3}). For this purpose, ultracold atoms in a three-dimensional optical lattice appear to be ideal, where the constant field is generated by gravity in the direction perpendicular to the two-dimensional layers. The field strength can be tuned by accelerating the optical lattice in the $\hat z$-direction, as already demonstrated in previous experiments witb one-dimensional structures.   
Furthermore, the transport of particles can be realized by preparing two particle reservoirs with different chemical potentials~\cite{krinner2016mapping}, for which the motion of particles can be traced and therefore the angular momentum discussed in Sec.~\ref{SecAng} can be measured. 
}

\hfill

\section*{Acknowledgement}
The authors acknowledge financial support from IBS (Project Code No. IBS-R024-D1). KWK is grateful to Carlo Danieli, Prof. J.E. Han, and Hee Chul Park for insightful discussions.

\appendix

\section*{Appendix A: Scattering matrix} 
Our calculations assume that the system-bath coupling is {sufficiently weak so} that its influence on the Wannier-Stark electronic structure is negligible.
Scattering between eigenmodes of the TWSL is mediated by incoherent scatterers such as background phonons and photons. {The scattering strengths for emission and absorption are given by}
\beq
    W_{nm} =& \rho(\omega) \left[1+n_{\omega} \right]  \int d^3 \vec x \,\nu_n(\vec x)\, \nu_m(\vec x) , \,\, &\text{(for  $\epsilon_n < \epsilon_m$)},\nonumber \\
    W_{nm} =& \rho(\omega) \left[n_{\omega} \right]  \int d^3 \vec x \,\nu_n(\vec x)\, \nu_m(\vec x) , \,\, &\text{(for  $\epsilon_n > \epsilon_m$)},\nonumber \\ \label{Wnm}
\eeq
where phonons absorb/emit the energy difference between two electronic eigenmodes $\omega=|\epsilon_m-\epsilon_n|$, $n_\omega = (e^{\omega/kT}-1)^{-1}$. The coupling strength between system and bath is $\rho(\omega) = c\omega^s e^{-\omega/\omega_D}$ where the phonon energy has a Debye frequency cutoff $\omega_D$. We take $s=1$ Ohmic dissipation.  The minimum energy of the phonon is bounded by the system size, $\omega_b=c/L_z$ and set to $\omega_b=0.01$ in our numerics.  $\nu_n(\vec x)=|\psi_n(\vec x)|^2$ is the local DoS of eigenmode $n$ obtained from the {diagonalisation of the} lattice model introduced in Eq.~\eqref{H}. 

To compute the  particle current across the system in Fig.~\ref{fig:1}(d), we introduce two thermalized reservoirs {with the same temperature as the thermal bath of TWSL}, localized at $z=\pm L_z/2$ with chemical potential $\mu=\pm V_0/2$, coupled to eigenmode $n$ in the system with strength $\mathcal A_n=a\sum_{x,y}|\psi_n(x,y,z=\pm L_z/2)|^2$ at energy $E=\epsilon_n$ with a constant $a$. Thus, the tunneling current, for example, to the reservoir 2 is computed as follows:
\beq
I = \sum_n \mathcal A_n^{\left(z\text{=}L_z/2\right)} \left( f(\epsilon_n) - \frac{1}{e^{(\epsilon_n+V_0/2)/kT}+1} \right),
\eeq
where $f(\epsilon_n)$ is determined {from} the steady state solution of the Pauli master equation in \eqref{pme}.


\section*{Appendix B: Topological  boundary modes vs. in-gap impurity states} \label{AppendixB}

\begin{figure*}
\begin{center}
\includegraphics[width=1.8\columnwidth]{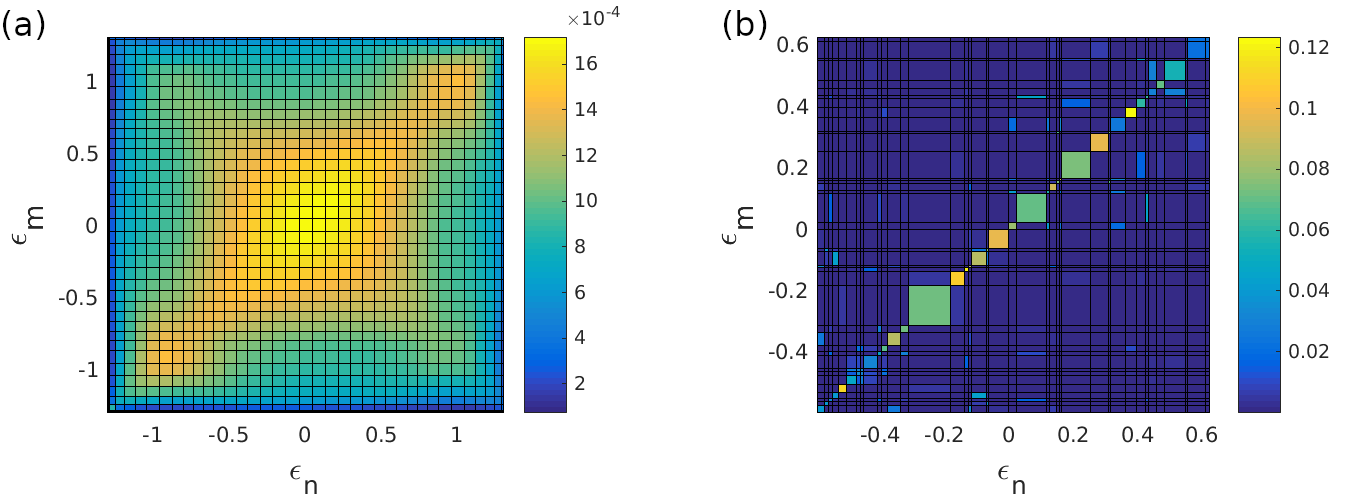}
    \caption{The overlap of the local DoS between modes $n$ and $m$, $B_{nm} = \int d^3\vec x\nu_m (\vec x)\nu_n(\vec x)$, is plotted for eigenmodes of model (i) (a) and model (ii) (b) in the energy gap. See Appendix B for details.  }
    \label{fig:B}
\end{center}
\end{figure*}

{The topological protection of boundary modes ensures the presence of eigenmodes continuously distributed in energy space which in turn leads to a smooth spatial  overlap of the local DoS among neighboring modes. Thus, the transport of particles by thermal relaxation remains stable. In spite of the significant reduction of the number of transport channels from the WS band ($\sim 2N_xN_y$) to the boundary modes ($\sim 4(N_x+N_y)$), the system is not insulating. In contrast, when a bulk energy gap is filled with impurity states, the transport of particles is governed by the scattering of particles from one impurity state to another. Because impurity states are spatially localized in the full three-dimensional space, the scattering strength is strongly reduced compared to that of topological boundary modes. Even worse, the scattering strength is exponentially decaying as a function of energy difference between two modes. Thus, the transport of particles through the energy gap filled with in-gap impurity states is  similar to an insulating phase. 

In this section we support the above argument by calculating the overlap of the local DoS between mode $n$ and $m$: 
\beq
B_{nm} = \int d^3\vec x \, |\psi_n(\vec x)|^2 |\psi_m(\vec x)|^2,
\eeq
which is a quantity directly related to the scattering strength in Eq.~\eqref{Wnm}. Note that when $n=m$, the above expression
becomes the inverse participation ratio, which is a meausure of spatial localization of eigenmodes. In the following we compute $B_{nm}$ for two models: (i) The model introduced in \eqref{H} with $F=12$ (see Fig.~\ref{fig:3} (a,c) for the related energy spectrum), and (ii) a model with the same set of microscopic parameters ($\mu, \alpha, \beta, J, \beta, F$) of the clean Hamiltonian but with periodic boundary conditions in the x- and y-directions so that edge/boundary modes are absent. In addition, for the second model we add an onsite impurity potential: 
\beq
H_{\text{imp}} &=& \sum_{n\in X} (V_n\sigma_0) c^\dg_n c_n,
\eeq
where the potential $V_n$  is added to a randomly chosen set of lattice sites $X=\{ n_1,n_2,\cdots n_M\}$ where $M = 2N_z(N_x+N_y)$, such as to maintain the same number of topological boundary modes as in model (i). The impurity potential is uniformly distributed betweeen $[-2\alpha, 2\alpha]$, where $\alpha=4$ is the hopping strength in xy-plane. In this way,  in-gap impurity states are introduced between the WS bands.

Figure~\ref{fig:B} (a) and (b) shows $B_{nm}$ for model (i) and (ii), respectively, for fourty eigenmodes near zero energy. First of all, the energy spacings (width betwen parallel lines) among neighboring topological boundary modes are remarkably uniform due to the level repulsion, while those of impurity modes are random as they are spatially localized in 3D space. Second, 
it follows that the overlap of the local DoS, $B_{nm}$, 
of model (i) is the largest for the nearest neighbors, $m=n+1$, and the value is roughly constant regardless of their eigenenergies. This provides advantages for the transport of particles through the topologically protected boundary modes, as the scattering strength is exponentially decaying with energy difference, $\omega = \epsilon_m-\epsilon_n$. In contrast,  $B_{nm}$ of model (ii)  between impurity modes does not show a conceivable correlation, indicating that the transport of particle through the impurity states is much less efficient. 

Lastly, particles in topological boundary modes in model (i) carry an angular momentum which linearly scales with the system size $L_{x,y}$ as discussed in Fig.~\ref{fig:5}. Instead, particles transporting through impurity modes do not carry a measurable angular momentum.  }

\bibliographystyle{apsrev4-1}
\bibliography{kun_biblio}

\hfill







\end{document}